# Factors Affecting the Precision of Electrostatic Computation of 3D MEMS Structures

N.Majumdar*, S.Mukhopadhyay
Saha Institute of Nuclear Physics,1/AF, Sector 1, Bidhannagar, Kolkata 700064, West Bengal, India
e-mail: nayana.majumdar@saha.ac.in

## Introduction

Of various types of Micro-Electro-Mechanical Systems (MEMS), a major class of MEMS is actuated by electrostatic force. Owing to their smallness these systems can be maneuvered by application of small voltages. For these devices, a precise electrostatic numerical simulation is necessary for the efficacy of their design and function. Usually the electrostatic analysis of MEMS design is carried out following Boundary Element Method (BEM). However, this method is inadequate for yielding accurate charge density on individual surfaces which is required to estimate the force acting on each plate. Several modified formulations have been developed to handle mathematical singularities arising out of the thinness of the plates and applied to estimate the surface charge densities and capacitance of a parallel plate capacitor [1].

Recently, a nearly exact Boundary Element Method (neBEM) solver has been developed [2,3] and used to solve difficult problems related to electrostatics of various devices [4,5]. The solver employs the analytic expression of potential and electric field obtained from the exact integration of Green's function due to a unit charge distribution distributed over a rectangular surface. Because of the exact analytic expressions, the neBEM has been found to yield very accurate results while solving critical problems which normally necessitate special formulations involving elegant, but difficult mathematics. In this work, the solver has been exercised over a wide range of thickness to length aspect ratio of a plate in a parallel plate capacitor relevant to the aspect ratio found in MEMS ($10^{-2} - 10^{-3}$). The effects of factors like discretization, adaptive mesh, exclusion of surfaces with small amount of charge on the charge density distribution as well as the capacitance have been studied. Since the force exerted on the MEMS finally depend on the charge distribution and its capacitance, its efficacy as both sensor and actuator is dependent critically on these parameters.

## Problem Definition

The geometry of the model has been illustrated in fig.1. The ratio of the gap between the plates to the length of the plate (d/L, d the gap and L the length) has been taken to be 0.2 while a range of aspect ratio of the plate thickness to length (h/L, h the thickness) $1 - 10^{-6}$ has been considered. The potentials applied to the upper and lower plates are V1 = 1 Volt and V2 = −1 Volt, respectively. It may be noted here that capacitance (C) has been computed as C = Q/2V where Q is the total charge on the upper plate while V is the potential on it. The value of the capacitance has been normalized by the standard parallel plate capacitance formula A ε /d where A is the area of the plate, d is the gap in between and ε is the permittivity of free space.

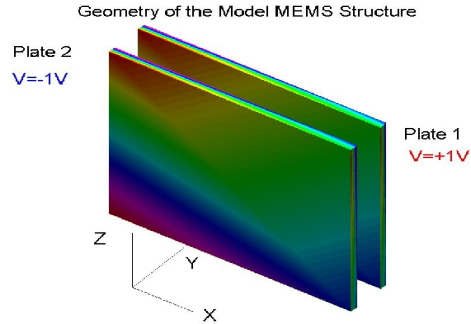

*Fig.1: Model of parallel plate capacitor*

### Results

a) Effect of discretization: The sharp rise of charge density at the edges of the plates can be reproduced more closely by refinement in the surface discretization. In fig.2, the charge distributions on the top and bottom surfaces of the upper plate with aspect ratio $10^{-3}$ (MEMS range) have been shown with different discretization schemes.

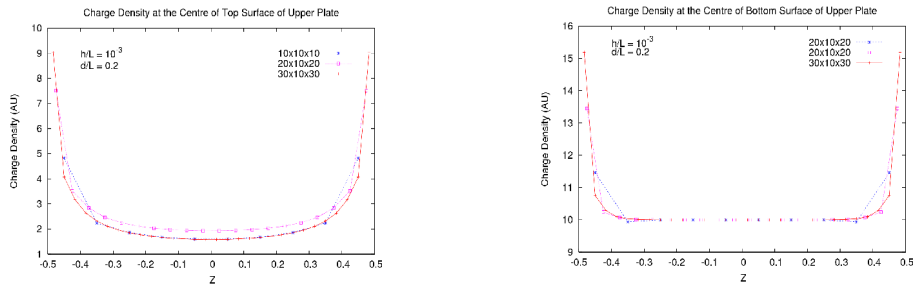

*Fig.2: Charge density distribution on top and bottom surfaces of the upper plate*

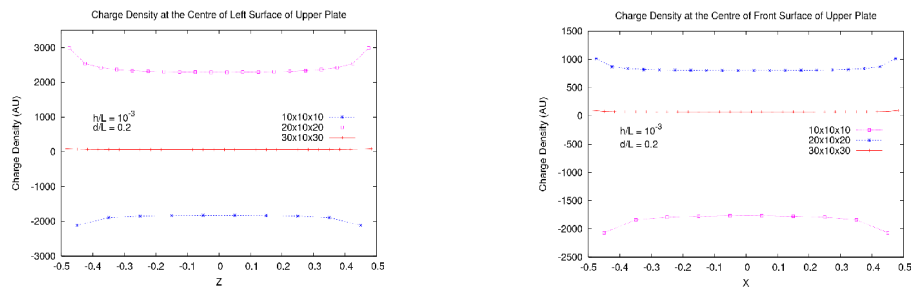

*Fig.3: Charge density distribution on left and front surfaces of the upper plate*

While it is true for top and bottom surfaces which are square in dimension and so the mesh elements, the result has been found different for the side surfaces where the mesh elements are not square, as shown in fig.3.

For smaller aspect ratio, this discretization scheme may not be appropriate since the charge density starts to diverge (see fig.4). It has also been observed that the aspect ratio of mesh elements on the sides does affect the accuracy of the charge distribution on the top and bottom surfaces. It is important to mention that the improper choice of discretization may lead to overestimation (in case of top surface) or underestimation (in case of bottom surface) of charge density as seen from fig.4. This inaccuracy may not be reflected in the capacitance of the system since it concerns to total charge only.

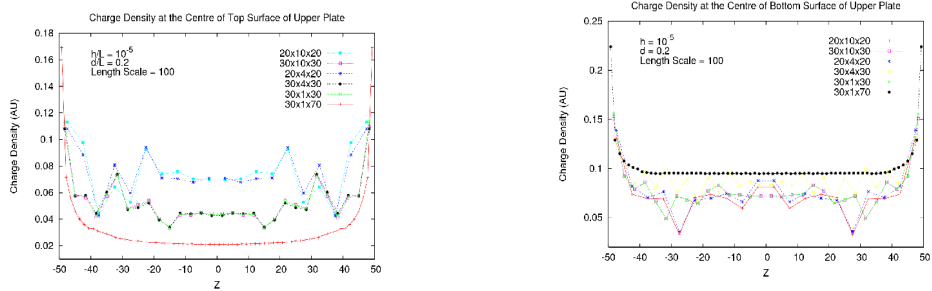

*Fig.4: Charge density distribution on top and bottom surfaces of upper plate*

Similar results have been obtained in case of side surfaces with above mentioned discretization scheme. The charge density has been better described by refined meshes with appropriate aspect ratio of mesh elements.

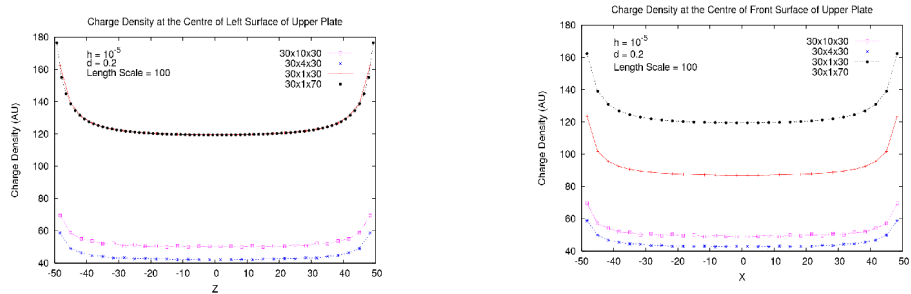

*Fig.5: Charge density distribution on left and front surfaces of upper plate*

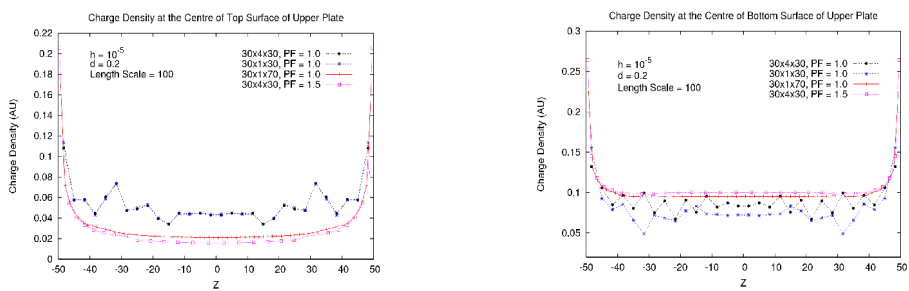

*Fig.6: Charge density distributions on top and bottom surfaces of upper plate with profile factor incorporated in discretization scheme*

b)Effect of profile factor: The profile of charge density can be better reproduced if an adaptive mesh is used instead of uniform mesh, also leading to a compromise between the extensive computation and accuracy of charge distribution. In fig.6, it has been demonstrated that the use of a small profile factor improves the results significantly even with a coarse discretization scheme.

c)Exclusion of smaller surfaces: In order to estimate the error introduced through the omission of the four side surfaces of the two plates shown in the geometry, estimated the capacitances for various values of h/L without considering the side plates and compared them with the complete geometry and other results [5]. The comparison has been presented in the following Table1.

Table 1: Variation of normalized capacitance for d/L = 0.2 with varying h/L

| h/L | Usual BEM [1] | Enhanced BEM [1] | Thin plate BEM [1] | Present (excluding side) | Present (including side) |
|---|---|---|---|---|---|
| 1.0 | 2.3975 | | | 1.259038 | 2.374961 |
| 0.1 | 3.3542 | 2.6631 | 1.2351 | 1.360813 | 1.757175 |
| 0.05 | | 1.7405 | 1.3879 | 1.392805 | 1.679710 |
| 0.01 | | 1.6899 | 1.5611 | 1.455047 | 1.590639 |
| 0.005 | | 1.6652 | 1.5874 | 1.475206 | 1.574417 |
| 0.001 | | 1.6221 | 1.6094 | 1.511291 | 1.558108 |
| 0.000001 | | | 1.6200 (1.5830 with finer mesh) | 1.539550 | 1.552190 |

From the table, it is easy to conclude that the omission of side plates leads to significant errors as long as h/L > 0.001. Below this limit, the side plates can be safely ignored since the total area of these surfaces remain negligible with respect to the top and bottom surfaces.

**Conclusions**

The nearly exact BEM (neBEM) solver has been used to estimate the effect of various factors affecting the accuracy of the numerical simulation of electrostatic properties of structures relevant to MEMS. The solver has been found to be accurate over a very wide range of aspect ratio because of the exact analytic nature of the foundation expressions. The study has shown that the choice of proper discretization is important in calculating electrostatic analysis of MEMS structures since it influences the convergence and profile of charge distribution on individual surfaces significantly. It may not be crucial in case of capacitance estimation which requires only the total charge.